\documentclass[10pt,twocolumn,twoside,submit]{JCNtran}
\usepackage{color}
\usepackage{amsmath}
\usepackage{amsfonts}
\usepackage{amssymb}
\usepackage{tabularx}
\usepackage{epsfig}
\usepackage{graphicx}
\usepackage[T1]{fontenc}
\usepackage{epstopdf}
\usepackage{multirow}
\usepackage{siunitx}
\usepackage[caption=true,font=footnotesize]{subfig}
\usepackage{fixltx2e}

\DeclareMathOperator*{\argmin}{arg\,min} 
\DeclareMathOperator*{\argmax}{arg\,max} 
\DeclareMathOperator*{\argmina}{argmin} 

\def\BibTeX{{\rm B\kern-.05em{\sc i\kern-.025em b}\kern-.08em
    T\kern-.1667em\lower.7ex\hbox{E}\kern-.125emX}}

\begin{document}

\title{Downlink Macro-diversity Precoding-aided Spatial Modulation}
\author{Manar Mohaisen and Vitalii Pruks
\thanks{Manuscript received January 31, 2017 and approved by Chungyoung Lee, Co-Editor-in-Chief, September 7, 2017.}
\thanks{This research was supported by Korea Tech research fund for the period of 2017-2018.}
\thanks{The authors are with the Department of Electrical, Electronics, and Communications Engineering, Korea Tech, Cheonan, R. of Korea, email: manar.subhi@koreatech.ac.kr}}
\markboth{JOURNAL OF COMMUNICATIONS AND NETWORKS, VOL. 17, NO. 4, AUGUST 2015}
{Mohaisen and Pruks: Downlink Macro-diversity Precoding-aided SM} \maketitle

\begin{abstract}
In this paper, a downlink macro-diversity precoding-aided spatial modulation (MD-PSM) scheme is proposed, in which two base stations (BSs) communicate simultaneously with a single mobile station (MS). As such, the proposed scheme achieved twice the spectral efficiency of the conventional PSM scheme. To render the demodulation possible, the two signal constellation sets used at the two BSs should be disjoint. Also, since the two BSs use the same spatial dimension, i.e., indices of receive antennas, the Minkowski sum of the two constellation sets should include unrepeated symbols. This is achieved through rotating the constellation set used by the second BS, where the error rate is also minimized. After obtaining the optimal rotation angles for several scenarios, a reduced complexity maximum-likelihood receiver is introduced. For an equal number of transmit and receive antennas of 4 and at a target BER of 10$^{-4}$, the simulation results show that the proposed MD-PSM scheme outperforms the conventional PSM by about 17.3 dB and 12.4 dB, while achieving the same and double the spectral efficiency, respectively. Also, due to the distributed nature of MD-PSM, it is shown that the diversity order of the novel MD-PSM scheme is twice that of the conventional PSM. 
\end{abstract}

\begin{keywords}
Spatial modulation, macro-diversity, multiple-input multiple-output (MIMO), MIMO precoding, maximum-likelihood detection.
\end{keywords}

\vspace{10pt}
\section{\uppercase{Introduction}}
\label{sec:introd}
\PARstart{S}{patial} modulation (SM) is a multiple-input multiple-output (MIMO) system in which both signal symbols, such as QAM/PSK symbols, and the indices of transmit antennas carry information from the transmitter to the receiver \cite{mesleh08}. Therefore, a terminal equipped with as few as a single RF chain and several physical antennas can enjoy a relatively high spectral efficiency as compared to the single-input single-output (SISO) case. In space-shift keying scheme, the conventional PSM/QAM modulation schemes are replaced by the presence or absence of energy \cite{jeg09}. That is, only the antenna whose index carries information transmits energy and the remaining antennas are turned off. Both schemes were generalized so that more than a single antenna can be simultaneously used. In this case, fewer physical antennas are required to achieve the same spectral efficiency as compared to the activation of a single antenna \cite{jeg08}-\cite{younis14}. In \cite{mesleh15}, a quadrature SM (QSM) scheme was proposed in which the real and imaginary parts of a signal constellation symbol, for instance $M$-ary PSK or QAM, are transmitted over an in-phase and quadrature spatial dimensions, respectively. The spatial constellation of the real and in-phase spatial dimensions are the sets of antennas' indices from which the real and imaginary parts of the signal symbol are transmitted. This increases the spectral efficiency by the number of bits transmitted over a single spatial dimension \cite{mesleh15}. Assuming QPSK modulation and four transmit antennas, SM transmits two bits over the signal symbols and two bits over a single spatial dimension, whereas QSM transmits four bits over two spatial dimensions. This increases the spectral efficiency from 4 bits/s/Hz to 6 bits/s/Hz \cite{mesleh15}. An extended SM (ESM) scheme was introduced in \cite{cheng15}, \cite{cheng16}, where information is carried not only by the signal and spatial symbols but also by the constellations transmitted from the antennas. It is shown that ESM scheme increases the number of transmitted bits per channel use by 1 or 2. Detailed comparison among these schemes is given in \cite{basar16}.

Precoding-aided spatial modulation (PSM) combines the conventional MIMO precoding techniques \cite{peel05} with spatial modulation, where a precoding vector is selected such that only a single, or multiple, intended receive antenna is activated and the energy received at the remaining antennas is minimized. The performance of the PSM scheme with linear zero-forcing (ZF) and minimum-mean square error (MMSE) precoders was addressed in \cite{yang11}. It is worth mentioning that more complex precoders can be used in combination with SM in order to achieve better error performance and diversity gains \cite{hochwald05}, \cite{mohaisen15}. Additionally, for an equal number of transmit and receive antennas, both SM and PSM achieve the same spectral efficiency.

In this paper, we propose a downlink macro-diversity PSM (MD-PSM) scheme, where two base stations (BSs) communicate simultaneously with a single mobile station (MS). The novel MD-PSK scheme, therefore, achieves twice the spectral efficiency of the conventional PSM scheme for the same system parameters. Since both BSs use the same spatial modulation set, i.e., the indices of the receive antennas, to convey information to the MS, it is probable that BSs' signals are sent to the same receive antenna. Therefore, the constellation set from which the non-zero, noiseless received signal is drawn is the union of the constellation sets used at both BSs and their Minkowski sum, i.e., element-wise addition of both sets. To render the demodulation possible at the receiver side, the following two conditions should be met: 1) The union of the two constellation sets used at the two BSs and their Minkowski sum, referred to as receive constellation set, should result in a set with unique symbols, even in the case when both BSs transmit the same number of bits per signal symbol, and 2) the minimum Euclidean distance between the symbols in the receive constellation set should be maximized in order to minimize the bit-error rate (BER). To achieve this, we propose to rotate the constellation set at the second BS, where the optimal rotation angle is obtained for several combinations of constellation sets. The rotation is performed by multiplying each symbol in the modulation set by $e^{j\theta}$, where $\theta$ is the rotation angle. This rotation preserves the power of symbols in the modulation set and the angles between them. A detailed analysis of the convergence of the rotation angle is also addressed, both using simulation results and analytically. Furthermore, the maximum-likelihood detector (MLD) was investigated and a reduced-complexity implementation, which benefits from the structure of the MD-PSM system, is proposed. 

The rest of this paper is as follows. In Section II, the conventional PSM scheme is reviewed. In Section III, the proposed MD-PSM scheme is introduced, a low-complexity MLD is advised, and the optimization of the rotation angle is addressed. In Section IV, simulation results and discussions of the findings are given. In Section V, we address the computational complexity of the MD-PSM, and we draw the paper's conclusions and introduce future works in Section VI.

\vspace{10pt}
\section{\uppercase{Related Work}}
\label{sec:sm}
In the considered system, a BS equipped with $n_T$ transmit antennas communicates with a MS equipped with $n_R = 2^{K}$ antennas. Only the BS has full knowledge of the channel state information (CSI). The channel matrix ${\mathbf{H}} \in {\mathbb{C}}^{n_R\times n_T} = [{\mathbf{h}}_1, \cdots, {\mathbf{h}}_{n_T}]$, where ${\mathbf{h}}_i$ is the $i$-th column of $\mathbf{H}$ and the element $h_{i,j}$ is the channel gain between the $i$-th receive and $j$-th transmit antenna. The channel gains are independent and identically distributed (i.i.d.) and follow a circularly-symmetric Gaussian distribution with mean and variance of zero and one, respectively. The precoding matrix ${\mathbf{P}} \in {\mathbb{C}}^{n_T\times n_R} = [{\mathbf{p}}_1, \cdots, {\mathbf{p}}_{n_R}]$ is designed according to performance optimization criteria, such as linear ZF and MMSE. The transmitted symbol $s \in \Omega$, where $|\Omega| = M = 2^q$, with $q$ denoting the number of bits per signal constellation symbol. At each transmission instant, a block of $m = (q + K)$ bits is transmitted. The first $q$ bits modulate a signal symbol and the remaining $K$ bits determine the index of the precoding vector ${\mathbf{p}}_i$ such that the transmitted vector of size $(n_T\times 1)$ is given by:
\begin{equation}
{\mathbf{x}} = \sqrt{\beta}{\mathbf{Pe}}_i s_k,
\label{eq:1}
\end{equation}
where ${\mathbf{e}}_i$ is the $i$-th column of the $(n_R\times n_R)$ identity matrix ${\mathbf{I}}_{n_R}$. Note that, in (\ref{eq:1}), both $i=1,\cdots,K$ and $k=1,\cdots,|\Omega|$ bear information. In the case of ZF precoding, the normalization factor $\beta = n_R/\text{Tr}({\mathbf{P}}{\mathbf{P}}^H)$, where $\mathbf{P}$ is the pseudo-inverse of $\mathbf{H}$ and $\text{Tr}(\mathbf{\cdot})$ is the trace operator. Accordingly, the received vector is given by:
\begin{equation}
{\mathbf{y}} = \sqrt{\beta}{\mathbf{HP}}{\mathbf{e}}_i s_k + {\mathbf{n}}.
\end{equation}
The noise vector ${\mathbf{n}}$ obeys the centered Gaussian distribution with covariance matrix ${\mathbb{E}}({\mathbf{n}}{\mathbf{n}}^H) = \sigma_n^2{\mathbf{I}}_{n_R}$, where $\sigma_n^2 = [(q+K)\gamma_b]^{-1}$ and $\gamma_b$ is the signal-to-noise ratio (SNR) per bit. Due to the zero structure of ${\mathbf{e}}_i$, the received signal at the $l$-th receive antenna in the case of ZF precoding is given by:
\begin{equation}
    y_l= 
\begin{cases}
    {\sqrt{\beta}}s_k + n_l,& \text{if } l=i\\        n_l,              & \text{otherwise}
\end{cases}
\end{equation}
Let ${\mathbf{r}} = {\mathbf{y}}/\sqrt{\beta}$, then the MLD of the PSM scheme is given by:
\begin{align}
(\hat{i}, \hat{s}_k) &= \argmin_{i,s_k} \|{\mathbf{r}} - {\mathbf{e}}_i s_k\|^2, \nonumber\\
										 &= \argmin_{i,s_k} \left|s_k\right|^2 - 2{\text{Re}}\{r_i^{*}s_k\}.
\label{eq:a}
\end{align}
where $\left|\cdot\right|$ and $\left\|\cdot\right\|$ are the absolute value of a real or complex scalar and vector norm, respectively.
\vspace{10pt}
\section{\uppercase{Macro-diversity PSM}}
\label{sec:mdpsm}
\subsection{System Model and Proposed MD-PSM}
\label{sec:mdpsm_sm}
A MIMO system is considered in which two BSs equipped with $n_{T_1}$ and $n_{T_2}$ transmit antennas, respectively, communicate simultaneously on the downlink with a single MS equipped with $n_R=2^K$ receive antennas, where $n_{T_1}, n_{T_2} \geq n_R$. Also, note that in recent deployments of massive MIMO systems, the number of antennas at the BS, $n_{T_i}$, for $i=1,2$, are much greater than the number of antennas at the MS. Let $s_{k_1} \in \Omega_a$ for $k_1=1,\cdots,M_1$ and $s_{k_2}^{\prime} \in \Omega_b$ for $k_2=1,\cdots,M_2$ be the signal symbols transmitted from the first and second BS, respectively, where $M_1=2^{q_1}$ and $M_2=2^{q_2}$. The signal constellation set $\Omega_a$ is a conventional QAM/PSK alphabet, while $\Omega_b$ is a rotated version of a conventional constellation set $\Omega$ such that $s_{k_2}^{\prime} \in \Omega_b = s_ke^{j\theta}$, where $s_k \in \Omega$ and $|\Omega_b| = |\Omega| = M_2$. The rotation angle $\theta$ is optimized so that the bit-error rate (BER) at the receiver is minimized. The optimization is addressed in the following Subsection. Now, let ${\mathbf{x}}_i$ and ${\mathbf{z}}_i$ be the precoded vector transmitted by the $i$-th BS and the corresponding noiseless received vector at the $n_R$ receive antennas, respectively. Then,
\begin{align}
{\mathbf{z}}_1 &= {\mathbf{H}}_1{\mathbf{x}}_1 = \sqrt{\beta_1}{\mathbf{H}}_1{\mathbf{P}}_1{\mathbf{e}}_{i_1} s_{k_1},\nonumber\\
{\mathbf{z}}_2 &= {\mathbf{H}}_2{\mathbf{x}}_2 = \sqrt{\beta_2}{\mathbf{H}}_2{\mathbf{P}}_2{\mathbf{e}}_{i_2} s_{k_2}^{\prime},
\label{eq:2}
\end{align}
where ${\mathbf{H}}_i$ and ${\mathbf{P}}_i$ are the matrix of channel gains from the $n_{T_i}$ transmit antennas to the $n_R$ receive antennas and the transmit precoding matrix employed at the $i$-th BS, respectively. In contrast to (\ref{eq:1}), $i_1, i_2, k_1$ and $k_2$ bear information, leading to a spectral efficiency of $(q_1+q_2+2K)$ bits/s/Hz, which is twice the spectral efficiency of the conventional PSM scheme for $q=q_1=q_2$. The $(n_R\times 1)$ received vector in the case of ZF precoding is therefore given by:
\begin{align} 
{\mathbf{y}} &= {\mathbf{z}} + {\mathbf{n}} = \left[
\begin{matrix}
{\mathbf{H}}_1 & {\mathbf{H}}_2
\end{matrix}
\right] 
\left[
\begin{matrix}
{\mathbf{x}}_1\\
{\mathbf{x}}_2
\end{matrix}
\right] + {\mathbf{n}},\nonumber\\
         &= \sqrt{\beta_1}{\mathbf{e}}_{i_1}s_{k_1} + \sqrt{\beta_2}{\mathbf{e}}_{i_2}s_{k_2}^{\prime} + {\mathbf{n}},
\label{eq:3}
\end{align}
where the noise covariance matrix ${\mathbb{E}}({\mathbf{n}}{\mathbf{n}}^H) = \sigma_n^2{\mathbf{I}}_{n_R}$, with $\sigma_n^2 = [(\bar{q}+K)\gamma_b]^{-1}$ and $\bar{q} = \frac{q_1+q_2}{2}$. Also, ${\mathbf{z}}$ is the noiseless received vector. In light of (\ref{eq:3}), we make the distinction between the following two cases:
\begin{itemize}
	\item $i_1 \neq i_2$: The received signal at the $l$-th antenna is given by:
	\begin{equation}
    y_l= 
\begin{cases}
    {\sqrt{\beta_1}}s_{k_1} + n_l,& \text{if } l=i_1\\
		{\sqrt{\beta_2}}s_{k_2}^{\prime} + n_l,& \text{if } l=i_2\\
		n_l,              & \text{otherwise}
\end{cases}
\label{eq:4}
\end{equation}
\item $i_1=i_2$: The received signal at the $l$-th antenna is given by:
	\begin{equation}
    y_l= 
\begin{cases}
    {\sqrt{\beta_1}}s_{k_1} + \sqrt{\beta_2}s_{k_2}^{\prime} + n_l,& \text{if } l=i_1=i_2\\
		n_l,              & \text{otherwise}
\end{cases}
\label{eq:5}
\end{equation}
\end{itemize}
Since the values of the normalization factors $\beta_1$ and $\beta_2$ are not equal, the optimization of the rotation angle should track the time-varying nature of $\beta_1$ and $\beta_2$, rendering the computational complexity of the optimization process high. Also, different values of $\beta$ renders the simplification of the ML receiver harder, as will be described in the next Subsection. It is shown in \cite{tulino04} that for $n_T>n_R$
\begin{equation}
{\mathbb{E}}\left[{\text{Tr}}\left(({\mathbf{HH}}^H)^{-1}\right)\right] = \frac{n_R}{n_T-n_R}.
\end{equation}
Therefore, the expected value of the normalization factor is simplified to
\begin{equation}
{\mathbb{E}}\left[\beta\right] = n_R/{\mathbb{E}}\left[{\text{Tr}}\left(({\mathbf{HH}}^H)^{-1}\right)\right] = n_T-n_R.
\end{equation}
As $n_T$ gets large, based on the law of large numbers ${\mathbb{E}}\left[\beta\right] \rightarrow n_T$ because $\left(({\mathbf{HH}}^H)^{-1}\right) \rightarrow \frac{1}{n_T}{\mathbf{I}}_{n_R}$ \cite{telatar99}. This simplification is valid for the case of transmissions of long duration. However, for short transmission duration, these simplifications miss the time-varying nature of $\beta$. Alternatively, we propose to use the average value of $\beta_1$ and $\beta_2$ as a unified normalization at both BSs. While this proposal does not affect the average value of the total transmission power from both BSs, the resulting noise amplification due to normalization, i.e., $1/\sqrt{\beta}$, has lower maximum value, mean value and variance, leading to better performance as compared to simply using $\beta_1$ and $\beta_2$ at the first and second BS, respectively. As a consequence, using the average value of the normalization factors at both BSs slightly increases the capacity of the MD-PSM system. In the sequel, a unified normalization factor of $\beta = \frac{\beta_1+\beta_2}{2}$ will be used. This exchange of information between the two BSs is made possible in the next generations of communication systems through backhauling \cite{ge14}. Let ${\mathbf{r}} = {\mathbf{y}}/\sqrt{\beta}$ and ${\mathbf{w}} = {\mathbf{n}}/\sqrt{\beta}$, (\ref{eq:4}) and (\ref{eq:5}) are rewritten as follows:
	\begin{equation}
    r_l= 
\begin{cases}
        s_{k_1} + w_l,& \text{if } l=i_1\\
		s_{k_2}^{\prime} + w_l,& \text{if } l=i_2\\
		w_l,              & \text{otherwise}
\end{cases}
\label{eq:6}
\end{equation}
in the case of $i_1 \neq i_2$, and
\begin{equation}
    r_l= 
\begin{cases}
      (s_{k_1} + s_{k_2}^{\prime}) + w_l =s_{k_3}^{\prime \prime} + w_l,& \text{if } l=i_1=i_2\\
	   w_l,              & \text{otherwise}
\end{cases}
\label{eq:7}
\end{equation}
in the case of $i_1 = i_2$. The symbol $s_{k_3}^{\prime \prime} \in \Omega_c$, where $\Omega_c = \Omega_a \oplus \Omega_b$, with $\oplus$ denoting the Minkowski sum and $|\Omega_c| = M_1M_2$ \cite{deberg}. Mathematically put, 
\begin{equation}
\Omega_c = \left\lbrace s_{k_1}+s_{k_2}^{\prime}|s_{k_1}\in \Omega_a, s_{k_2}^{\prime}\in \Omega_b\right\rbrace.
\end{equation}
Discarding the noise term in (\ref{eq:6}) and (\ref{eq:7}), the non-zero, noiseless elements of the received vector, denoted by $m_l$, belong to the constellation set $\Omega_d = \Omega_a\cup\Omega_b\cup\Omega_c$, where $\cup$ is the set union operator and $|\Omega_d| = (M_1+M_2+M_1M_2)$. More precisely,
\begin{equation}
    m_l= 
\begin{cases}
        s_{k_1},& \text{if } i_1\neq i_2, l=i_1, k_1 = \{1,\cdots,M_1\}\\
		s_{k_2}^{\prime},& \text{if } i_1\neq i_2, l=i_2, k_2 = \{1,\cdots,M_2\}\\
		s_{k_3}^{\prime \prime},& \text{if } l=i_1=i_2, k_3 = \{1,\cdots,M_1M_2\}
\end{cases}
\label{eq:6x}
\end{equation}

\subsection{Maximum-likelihood Detector}
\label{sec:mld}
The receiver employs the maximum-likelihood principle to estimate the signal and spatial symbols as follows:
\begin{align}
(\hat{i}_1, \hat{i}_2, \hat{s}_{k_1}, \hat{s}_{k_2}^{\prime}) &= \argmina_{i_1, i_2, s_{k_1}, s_{k_2}^{\prime}}\| {\mathbf{r}} -  {\mathbf{e}}_{i_1}s_{k_1} - {\mathbf{e}}_{i_2}s_{k_2}^{\prime}\|^2,\nonumber \\
		&= \argmina_{i_1, i_2, s_{k_1}, s_{k_2}^{\prime}} \|{\mathbf{e}}_{i_1}s_{k_1} + {\mathbf{e}}_{i_2}s_{k_2}^{\prime}\|^2 \nonumber\\
		&\hspace{45pt}-2{\text{Re}}\{r_{i_1}^{*}s_{k_1} + r_{i_2}^{*}s_{k_2}^{\prime}\}
\label{eq:9}
\end{align}
Due to the zero structure of ${\mathbf{e}}_{i_1}$ and ${\mathbf{e}}_{i_2}$, (\ref{eq:9}) is rewritten as follows:
\begin{align}
&(\hat{i}_1, \hat{i}_2, \hat{s}_{k_1}, \hat{s}_{k_2}^{\prime}) = \argmina_{i_1, i_2, s_{k_1}, s_{k_2}^{\prime}}\nonumber\\ 
&\begin{cases}
    \frac{\mid s_{k_1}\mid^2 + \mid s_{k_2}^{\prime}\mid^2}{2} - \text{Re}\{r_{i_1}^{*}s_{k_1} + r_{i_2}^{*}s_{k_2}^{\prime}\},& \text{if } i_1\neq i_2\\
		\frac{\mid s_{k_3}^{\prime \prime}\mid^2}{2} - \text{Re}\{r_{i_1}^{*}s_{k_3}^{\prime\prime}\},& \text{if } i_1=i_2
\end{cases}
\label{eq:10}
\end{align}
Since the first terms in (\ref{eq:10}), i.e., $\frac{\mid s_{k_1}\mid^2 + \mid s_{k_2}^{\prime}\mid^2}{2}$ and $\frac{\mid s_{k_3}^{\prime \prime}\mid^2}{2}$, are deterministic, they can be precomputed and stored in memory. The first term in (\ref{eq:10}) is divided by 2 in order to avoid multiplying the second term by 2 at each evaluation. This reduces the computational complexity by $(n_R-1)(M_1+M_2)$ real multiplications. The first line of (\ref{eq:10}), i.e., the case when $(i_1 \neq i_2)$, can be split into two separate optimization problems as follows:
\begin{align}
(\hat{i}_1, \hat{s}_{k_1}) &= \argmina_{i_1,s_{k_1}} \frac{\mid s_{k_1}\mid^2}{2} - \text{Re}\{r_{i_1}^{*}s_{k_1}\},\nonumber\\
(\hat{i}_2, \hat{s}_{k_2}^{\prime}) &= \argmina_{i_2, s_{k_2}^{\prime}} \frac{\mid s_{k_2}^{\prime}\mid^2}{2} - \text{Re}\{r_{i_2}^{*}s_{k_2}^{\prime}\}.
\label{eq:15}
\end{align}
This is rendered possible because the cost functions in the first and second line of (\ref{eq:15}) are independent. Equation (\ref{eq:15}) is further simplified to 
\begin{align}
(\hat{i}_1, \hat{s}_{k_1}) &= \argmax_{i_1,s_{k_1}} \text{Re}\{r_{i_1}^{*}s_{k_1}\},\nonumber\\
(\hat{i}_2, \hat{s}_{k_2}^{\prime}) &= \argmax_{i_2, s_{k_2}^{\prime}} \text{Re}\{r_{i_2}^{*}s_{k_2}^{\prime}\},
\label{eq:16}
\end{align}
in the case of $M$-ary phase-shift keying (M-PSK) modulation schemes, where the conveyed information is transmitted over the phase of the signal symbols. In this case, the signal symbols have equal square Euclidean norm.

Given the constellation sets $\Omega_a$ and $\Omega_b$, $\frac{\mid s_{k_1}\mid^2}{2}$ and $\frac{\mid s_{k_2}^{\prime}\mid^2}{2}$ are easily computed. The computation of $\frac{\mid s_{k_3}^{\prime \prime}\mid^2}{2}$, where $s_{k_3}^{\prime \prime}\in \Omega_c$, however, depends on the modulation schemes used at both BSs and the value of the optimal rotation angle, $\theta_\text{opt}$. Further characterization of the symbols of $\Omega_c$ is given in the following. We make the distinction between 2 cases: 1) both BSs use PSK modulation schemes, and 2) either or both of BSs use QAM schemes with $M > 4$. In the sequel, the QPSK modulation set used through this paper is given by $\{1, j, -1, -j\}$. This choice was made in order to emphasize on the fact that lower-order PSK modulation sets are geometrically subsets of higher-order ones. The results for the first and second case are depicted in Table \ref{tab:1} and \ref{tab:2}, respectively. The first case is explained in the following. Let $\Omega_c$ be split into $N$ disjoint subsets such that $\Omega_c = \cup_{i=1}^{N} \Omega_i$, where the symbols belonging to each of these subsets have an equal Euclidean norm. Let $\phi_i$ be the phase of the first symbol in the subset $\Omega_i$, then the symbols of the subset are uniformly distributed over a circle and located at angles 
\begin{equation}
\phi_i + \frac{2\pi(l-1)}{|\Omega_i|},\,\,\,l=1,\cdots, |\Omega_i|,
\end{equation}
where $\phi_i$ is given in Table \ref{tab:1} and $|\Omega_i|=|\Omega_a|$. For example, $|\Omega_i|=$ 2, 4 and 8, in the case of BPSK-8PSK, QPSK-8PSK, and 8PSK-8PSK, respectively, and the number of subsets, $N$, is equal to $|\Omega_b|$.
\begin{table}
\small
\centering
\caption{The values $|s_k^{\prime \prime}(\theta)|^2/2$ and the corresponding $\phi_i$ when both base stations use PSK modulation schemes.}
\label{table:pskpsk}
\def\arraystretch{1.4}

\begin{tabular}{l|l|l|l}
  \hline
  Modulation & $|s_k^{\prime \prime}(\theta)|^2/2$ & $\phi_i$ & $\theta_{\text{opt}}$\\
  \hline
	\multirow{2}{*}{BPSK-BPSK} & $1+\cos(\theta)$ & $\frac{\theta}{2}$& \multirow{2}{*}{$\geq 60\si{\degree}$}\\
                           & $1-\cos(\theta)$ & $\frac{\theta}{2}+\frac{\pi}{2}$\\
\hline
\multirow{4}{*}{\begin{tabular}{@{}c@{}}BPSK-QPSK$/$ \\ QPSK-QPSK\end{tabular}} & $1+\cos(\theta)$ & $\frac{\theta}{2}$& \multirow{4}{*}{$30\si{\degree}$}\\
                           & $1-\cos(\theta)$ & $\frac{\theta}{2}+\frac{\pi}{2}$\\
													 & $1+\sin(\theta)$ & $\frac{\theta}{2}-\frac{\pi}{4}$\\
													 & $1-\sin(\theta)$ & $\frac{\theta}{2}+\frac{\pi}{4}$\\
\hline
\multirow{8}{*}{\begin{tabular}{@{}c@{}}BPSK-8PSK$/$ \\ QPSK-8PSK$/$ \\8PSK-8PSK\end{tabular}} & $1+\cos(\theta)$ & $\frac{\theta}{2}$ & \multirow{8}{*}{\begin{tabular}{@{}c@{}}$15\si{\degree}, 30\si{\degree}/$ \\ $15\si{\degree}, 30\si{\degree}/$ \\$17.3\si{\degree}, 27.7\si{\degree}$\end{tabular}}\\
													 & $1-\cos(\theta)$ & $\frac{\theta}{2}+\frac{\pi}{2}$\\
 													 & $1+\sin(\theta)$ & $\frac{\theta}{2}-\frac{\pi}{4}$\\
													 & $1-\sin(\theta)$ & $\frac{\theta}{2}+\frac{\pi}{4}$\\
                           & $1+\cos(\theta+\frac{\pi}{4})$ & $\frac{\theta}{2}+\frac{\pi}{8}$ \\
													 & $1-\cos(\theta+\frac{\pi}{4})$ & $\frac{\theta}{2}-\frac{3\pi}{8}$\\
												   & $1+\sin(\theta+\frac{\pi}{4})$ & $\frac{\theta}{2}-\frac{\pi}{8}$ \\
													 & $1-\sin(\theta+\frac{\pi}{4})$ & $\frac{\theta}{2}+\frac{3\pi}{8}$ \\
													 
\hline
 \end{tabular}

\label{tab:1}
 \end{table}


\begin{table}
\small
\centering
\caption{The values $|s_k^{\prime \prime}(\theta)|^2/2$ when one base station uses 16QAM and the second uses any of the PSK modulation schemes.}
\label{table:pskqam}
\def\arraystretch{1.4}

\begin{tabular}{l|l|l}
  \hline
  Modulation & $|s_k^{\prime \prime}(\theta)|^2/2$ & $\theta_{\text{opt}}$\\
  \hline
\multirow{16}{*}{\begin{tabular}{@{}c@{}}BPSK-16QAM$/$ \\ QPSK-16QAM$/$ \\8PSK-16QAM\end{tabular}} & $1.4+\sqrt{1.8}\cos(\theta+\frac{\pi}{4})$ & \multirow{16}{*}{\begin{tabular}{@{}c@{}}$32.1\si{\degree}/$ \\ $32.1\si{\degree}/$ \\$8.4\si{\degree}, 36.6\si{\degree}$\end{tabular}}\\
                           & $1.4-\sqrt{1.8}\cos(\theta-\frac{\pi}{4})$\\
													 & $1.4-\sqrt{1.8}\cos(\theta+\frac{\pi}{4})$ \\
													 & $1.4+\sqrt{1.8}\cos(\theta-\frac{\pi}{4})$\\
													 & $0.6+\sqrt{0.2}\cos(\theta+\frac{\pi}{4})$\\
                           & $0.6-\sqrt{0.2}\cos(\theta-\frac{\pi}{4})$\\
													 & $0.6-\sqrt{0.2}\cos(\theta+\frac{\pi}{4})$\\
													 & $0.6+\sqrt{0.2}\cos(\theta-\frac{\pi}{4})$\\
													 & $1+\cos(\theta+\alpha)$\\
													 & $1-\sin(\theta+\alpha)$\\
													 & $1-\cos(\theta+\alpha)$\\
													 & $1+\sin(\theta+\alpha)$\\
													 & $1-\sin(\theta-\alpha)$\\
													 & $1-\cos(\theta-\alpha)$\\
													 & $1+\sin(\theta-\alpha)$\\
													 & $1+\cos(\theta-\alpha)$\\
\hline
 \end{tabular}
\label{tab:2}
 \end{table}

Table \ref{tab:1} summarizes these results as a function of the rotation angle $\theta$ for several scenarios, where both BSs use $M$-ary PSK modulation schemes. The leading 1 in the second column of Table \ref{tab:1} can be neglected without affecting the optimization problem, leading to further reduction in the computational complexity. For instance, in the case of QPSK-QPSK, where both BSs use QPSK modulation, the pre-computation of $|s_k^{\prime \prime}(\theta)|^2/2$ requires the evaluation of {\it{only}} $\cos(\theta_{\text{opt}})$ and $\sin(\theta_{\text{opt}})$, which can also be precomputed and saved in memory.

In the second case, i.e., at least one of the BSs uses 16QAM, the 16QAM modulation set is split into four disjoint and scaled 4QAM subsets defined by the following pairs of Euclidean norm and the angle of the first symbol of each subset.
\begin{align}
&\{(\sqrt{0.2}, \frac{\pi}{4}), (\sqrt{1.8}, \frac{\pi}{4}),(1, \alpha), (1, \frac{\pi}{2}-\alpha)\},
\label{eq:20}
\end{align}
where $\alpha=\arctan(1/3)\approx 18.435\si{\degree}$. Table \ref{tab:2} shows the obtained values of $|s_k^{\prime \prime}(\theta)|^2/2$ when the first BS uses $M$-ary PSK and the second uses 16QAM scheme. It is worth recalling that each of the values listed in the second column of Table \ref{tab:2} is repeated $M$ times as the second BS employs $M$-ary PSK modulation. We dropped the computation of $\phi_i$ since a closed-form formula is hard to obtain. Also, when both BSs use 16QAM, we obtained 40 different values for $|s_k^{\prime \prime}(\theta)|^2/2$, each repeated 4 or 8 times, resulting in $M_1M_2 = 256$ values. These values, which are omitted due to space limits, can be obtained through careful consideration of the subsets given in (\ref{eq:20}).

Based on (\ref{eq:15}) and regardless of whether $\Omega = \Omega_a$ or not, $\Omega_a \cup \Omega_b$ must be a set; that is, $s_{k_1} \in \Omega_a \neq s_{k_2}^{\prime} \in \Omega_b , \forall k_1\in \{1,\cdots,M_1\} \text{ and } k_2\in \{1,\cdots,M_2\}.$ This is the {\it{first condition}} that the choice of the optimal angle $\theta_{\text{opt}}$ should fulfil. The {\it{second condition}} is derived from the second line of (\ref{eq:10}): the symbols $s_{k_3}^{\prime \prime} \in \Omega_c$ for $k_3 \in \{1,\cdots,M_1M_2\}$ should be distinct in order to render the demodulation at the receiver side feasible. In other words, the mapping from $\Omega_a$ and $\Omega_b$ to $\Omega_c$ must be unique. The first and second conditions are applied to exclude certain rotation angles which violate their basic concept of uniqueness. However, an infinite set of potential rotation angles fulfil both conditions. In the following Subsection, we introduce a strict condition that  embodies both the first and second conditions and minimizes the BER.
\begin{figure}
\centering
\subfloat[]{\label{subfig:a} \includegraphics[scale=0.7]{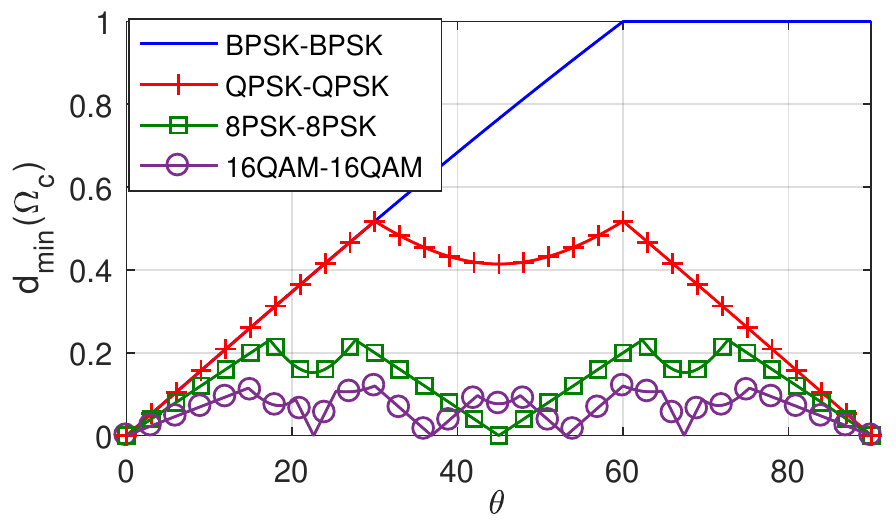}}\\
\subfloat[]{\label{subfig:b} \includegraphics[scale=0.7]{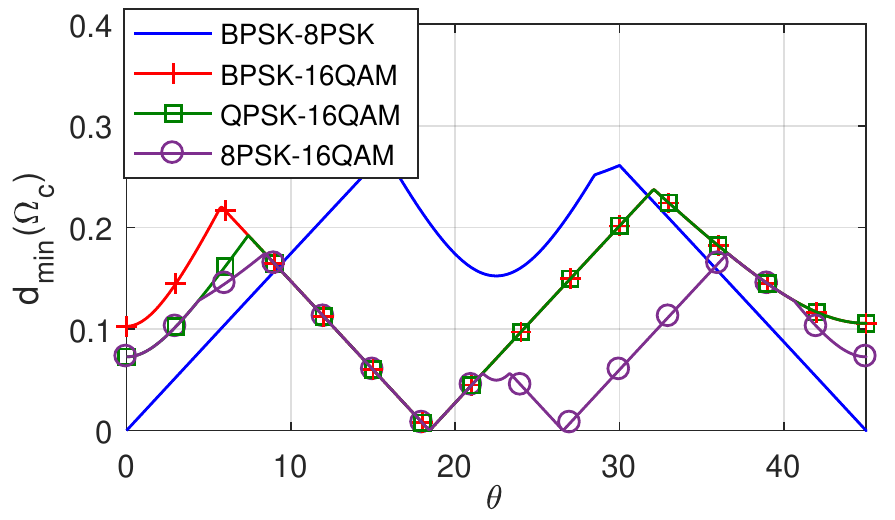}}
\caption{Minimum Euclidean distance for several combinations of modulation schemes versus the rotation angle $\theta$, with BSs using a) same modulation schemes, and b) different modulation schemes.}
\label{fig:med}
\end{figure}
\begin{figure*}
\centering
\includegraphics[scale=0.7]{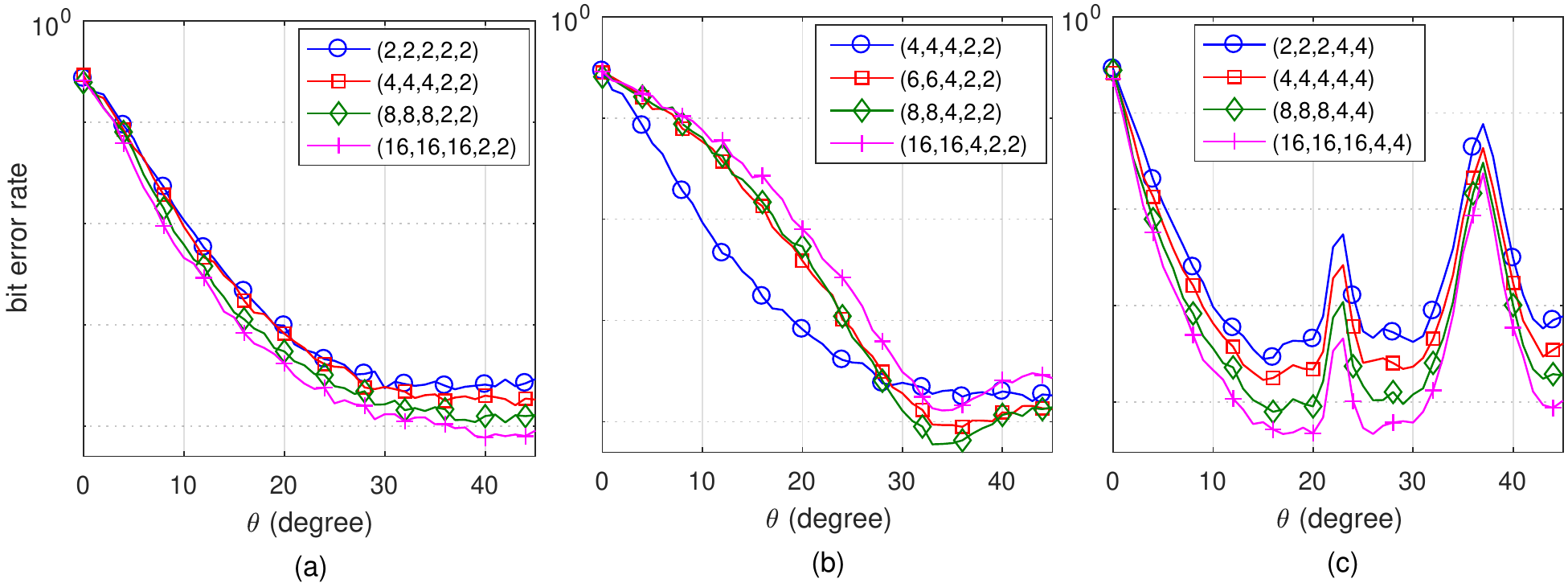}
\caption{BER performance of the proposed MD-PSM versus the rotation angle $\theta$ in the range [0,$\pi/4$] for a) QPSK and $n_{T_i}=n_{R}$, b) QPSK and $n_{T_i} \geq n_{R}$, and c) 16QAM and QPSK and $n_{T_i}=n_{R}$.}
\label{fig:angle_sim}
\end{figure*}
\subsection{Optimization of the Rotation Angle}
\label{sec:rot}
As described in (\ref{eq:6}) and (\ref{eq:7}), the system equation reduces to the additive-white Gaussian noise model. The minimum Euclidean distance between the signal symbols of the receive constellation set $\Omega_d = \Omega_a\cup \Omega_b \cup \Omega_c$, denoted by $d_{\text{min}}(\Omega_d)$, is a determining factor of the BER performance: a larger $d_{\text{min}}(\Omega_d)$ leads to smaller BER. Let $m_i, m_k \in \Omega_d$, then the optimal rotation angle is given by:
\begin{equation}
{\hat{\theta}} = \argmax_{\theta} \left\{\min_{m_i, m_k, i\neq k} \left|m_i-m_k\right|^2\right\},
\label{eq:20x}
\end{equation}
The rotation angle is optimized through simulations and results are depicted in Fig. \ref{fig:med}(a) and (b) for several combinations of modulation schemes used at the two BSs. For instance, the legend BPSK-16QAM refers to the scenario where the first and second BS uses BPSK and 16QAM schemes, respectively, and the rotation is applied to the constellation set at the second BS. The curves for BPSK-QPSK and QPSK-QPSK, BPSK-8PSK and QPSK-8PSK are equivalent. Except of the case of BPSK-BPSK, the curves of $d_{\text{min}}(\Omega_d)$ are even symmetric around the angle $\theta = \pi/4$ (rad). Note that the optimal rotation angle when both BSs use 16QAM is $30\si{\degree}$. The optimal rotation angles for the other cases are listed in the last columns of Tables \ref{tab:1} and \ref{tab:2}. 

While (\ref{eq:20x}) obtains an accurate value of $d_{\text{min}}(\Omega_d)$, it does not take into consideration the probabilities of symbols in the receive constellation set. These probabilities will affect the {\it{expected value}} of the Euclidean distance among signal symbols in $\Omega_d$. More precisely, the BER performance of the ML receiver depends on the average value of the Euclidean distance, which is defined as
\begin{equation}
d_{i,k} = \left|{\mathbf{s}}_i-{\mathbf{s}}_k\right|^2,
\end{equation}
where ${\mathbf{s}}_i$ and ${\mathbf{s}}_k$ are the vectors transmitted by the first and second BS, respectively. As such, the average Euclidean distance depends on the signal symbols in $\Omega_d$. Based on (\ref{eq:6}) and (\ref{eq:7}), the probabilities of the noiseless, non-zero received symbols $m_k\in \Omega_d$ are given as follows:

\begin{align}
{\text{Pr}[m_k \in \Omega_d \cap m_k \in \Omega_a]} &= {\text{Pr}[m_k \in \Omega_d \cap m_k \in \Omega_b]}, \nonumber\\
                                                    &= \sum_{k=1}^{M_1}{\text{Pr}[m_{k}\in \Omega_a]}\nonumber\\
																										&= \sum_{k=1}^{M_2}{\text{Pr}[m_{k}\in \Omega_b]},\nonumber\\
                                                    &= {\text{Pr}}[i_1 \neq i_2] = 1-\frac{1}{n_R},\nonumber\\
{\text{Pr}[m_k \in \Omega_d \cap m_k \in \Omega_c]} &= \sum_{k=1}^{M_1M_2}{\text{Pr}[m_{k}\in \Omega_c]}\nonumber\\
                                                    &= {\text{Pr}}[i_1 = i_2] = \frac{1}{n_R}.
\label{eq:30x}
\end{align}
The events in the first line of (\ref{eq:30x}) occur simultaneously. This implies that as $n_R$ grows large, the probability that the transmit symbol $\in \Omega_c$ vanishes. In this case, the optimal rotation angle will converge to the one where $\Omega_d = \Omega_a\cup\Omega_b$. As such, the minimum Euclidean distance is increased, which leads to better BER performance. Since the signal symbols in each of the sets $\Omega_a$, $\Omega_b$, and $\Omega_c$ are equiprobable, then
\begin{align}
{\text{Pr}[m_{k}=s_{k_1}\in \Omega_a]} &= \frac{n_R-1}{M_1n_R},\,\,\,k_1=1,\cdots,M_1,\nonumber\\
{\text{Pr}[m_{k}=s_{k_2}^{\prime}\in \Omega_b]} &= \frac{n_R-1}{M_2n_R},\,\,\,k_2=1,\cdots,M_2,\nonumber\\
{\text{Pr}[m_{k}=s_{k_3}^{\prime \prime}\in \Omega_c]} &= \frac{1}{M_1M_2n_R},\,\,\,k_3=1,\cdots,M_1M_2.
\end{align}

Based on (\ref{eq:30x}) and the above discussion, it stems out that
\begin{align}
&\lim_{n_R\rightarrow \infty}{\text{Pr}[m_k \in \Omega_d \cap m_k \in \Omega_a]} = \nonumber\\
&\lim_{n_R\rightarrow \infty}{\text{Pr}[m_k \in \Omega_d \cap m_k \in \Omega_b]} \approx 1,\nonumber\\
&\lim_{n_R\rightarrow \infty}{\text{Pr}[m_k \in \Omega_d \cap m_k \in \Omega_c]} \approx 0.
\label{eq:40x}
\end{align}
This implies that while $d_{\text{min}}(\Omega_d)$ is fixed for given $\Omega_a$ and $\Omega_b$, the {\it{average value}} of the Euclidean distance of $\Omega_d$ converges to $d_{\text{min}}(\Omega_a \cup \Omega_b)$ since ${\text{Pr}[m_{k}\in \Omega_c]} \approx 0$. For instance, when both BSs use QPSK, $\theta_{\text{opt}}$ converges to $\pi/4$ as $n_R$ grows large.

\vspace{10pt}
\section{\uppercase{Simulation Results and Discussions}}
\label{sec:simulations}
Each BS has perfect knowledge of the state information of the channel coupling its transmit antennas to the receive antennas of the single MS, whereas the MS does not have this information. We assume that BSs exchange their computed normalization factors, $\beta_1$ and $\beta_2$, and both BSs use a unified normalization factor of $\frac{\beta_1+\beta_2}{2}$. In the legends of the following figures, the conventional PSM is indicated by the three-element tuple $(n_T, n_R, q)$, while the proposed scheme is indicated by the five-element tuple $(n_{T_1}, n_{T_2}, n_{R}, q_1, q_2)$.

Before evaluating the performance of the proposed MD-PSM scheme, the choice of the optimal rotation angle is addressed through Monte-Carlo simulations. Figure \ref{fig:angle_sim} depicts the BER performance of the novel scheme for several scenarios. Figure \ref{fig:angle_sim}(a) depicts the BER performance of the MD-PSM scheme for several values of $n_{T_i} = n_R$ with QPSK modulation. The optimal rotation angle $\theta_{\text{opt}}$ is equal to $30\si{\degree}$, $33\si{\degree}$, $37\si{\degree}$ and $40\si{\degree}$ in the case of $n_R=2$, 4, 8 and 16, respectively. To emphasize on this finding, Fig. \ref{fig:angle_sim}(b) depicts the BER performance for a fixed $n_R$ and several values of $n_{T_1}=n_{T_2} \geq n_R$, where the noise effect is therefore reduced through increasing the diversity order and the minimum singular values of $\textbf{H}_1$ and $\textbf{H}_2$. The rotation angle in this case is clearly equal to $\theta_{\text{opt}} = 33\si{\degree}$. Figure \ref{fig:angle_sim}(c) depicts the performance of the proposed scheme when both BSs use 16QAM. For $n_R=2,4,8$ and $16$, $\theta_{\text{opt}} = 15\si{\degree}, 15\si{\degree}, 16\si{\degree}$ and $17\si{\degree}$, respectively. We also remark that the BER performance is almost the same in the case of $n_{T_i}=n_R=16$ at $\theta = 17\si{\degree}$ and $26\si{\degree}$. Since $d_{\text{min}}(\Omega_a \cup \Omega_b) = 25.5\si{\degree}$, which is obtained through simulations, these results coincide with our remarks on the convergence of $\theta_\text{opt}$, given in Section \ref{sec:rot}. 

Offline optimization of the rotation angle is achievable with low computational complexity because the search is performed over a limited range of angles that depends on the modulation order. For instance, the search range in the case of QPSK and 16PSK are [30, 45] and [9.7, 11.25], respectively. This range is narrowed as the modulation order is increased, and the computational complexity of the search problem is therefore reduced. The lower and upper bound of these intervals is the optimal rotation angle considering equiprobable symbols in the set $\Omega_d$ and when $n_R$ goes to infinity, respectively.

\begin{figure}
\centering
\includegraphics[scale=0.82]{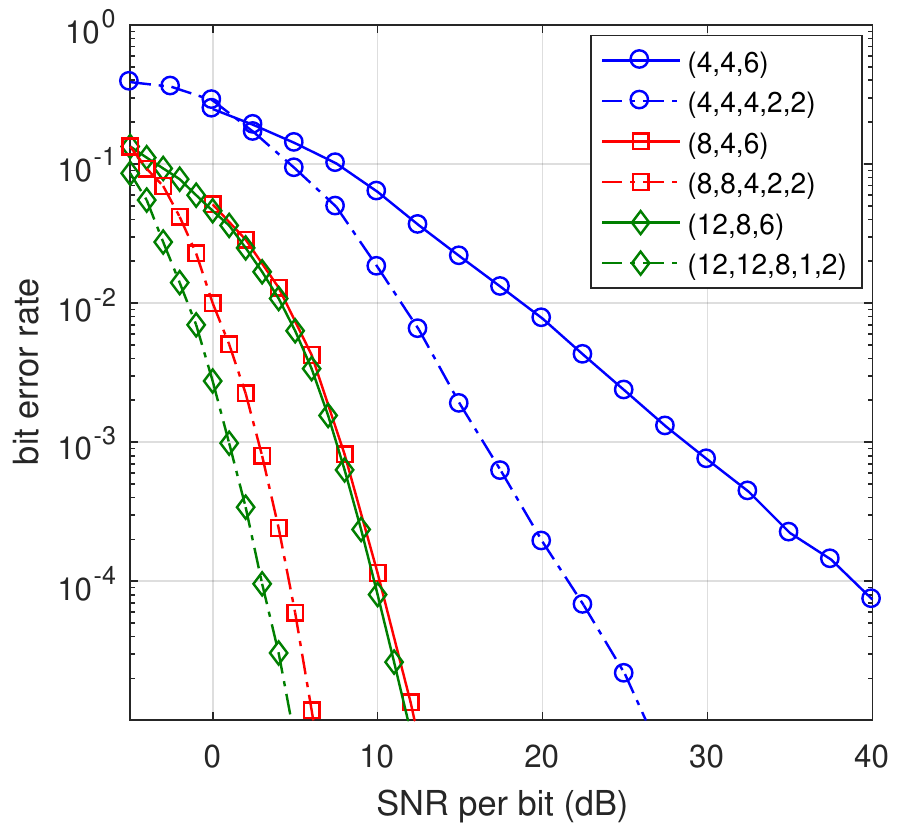}
\caption{BER performance of the proposed MD-PSM versus that of PSM for the same spectral efficiency.}
\label{fig:sim_samese}
\end{figure}

Figure \ref{fig:sim_samese} depicts the performance of the proposed MD-PSM versus that of the conventional PSM for the same spectral efficiency. The modulation order is adjusted so that both MD-PSM and PSM schemes achieve the same spectral efficiency while keeping $n_T=n_{T_1}=n_{T_2}$ for the same $n_R$. In the first case, where $n_T=n_{T_1}=n_{T_2} = n_R=4$, the proposed scheme outperforms PSM by more than 17.3 dB at a target BER of $10^{-4}$. In the case of $n_T=n_{T_1}=n_{T_2}=8$ and $n_R=4$, the proposed scheme outperforms the conventional PSM by 5.5 dB. This gain increases to 6.8 dB in the case of $n_T=n_{T_1}=n_{T_2}=12$ and $n_R=8$, all at a target BER of $10^{-4}$. 
\begin{figure}
\centering
\includegraphics[scale=0.82]{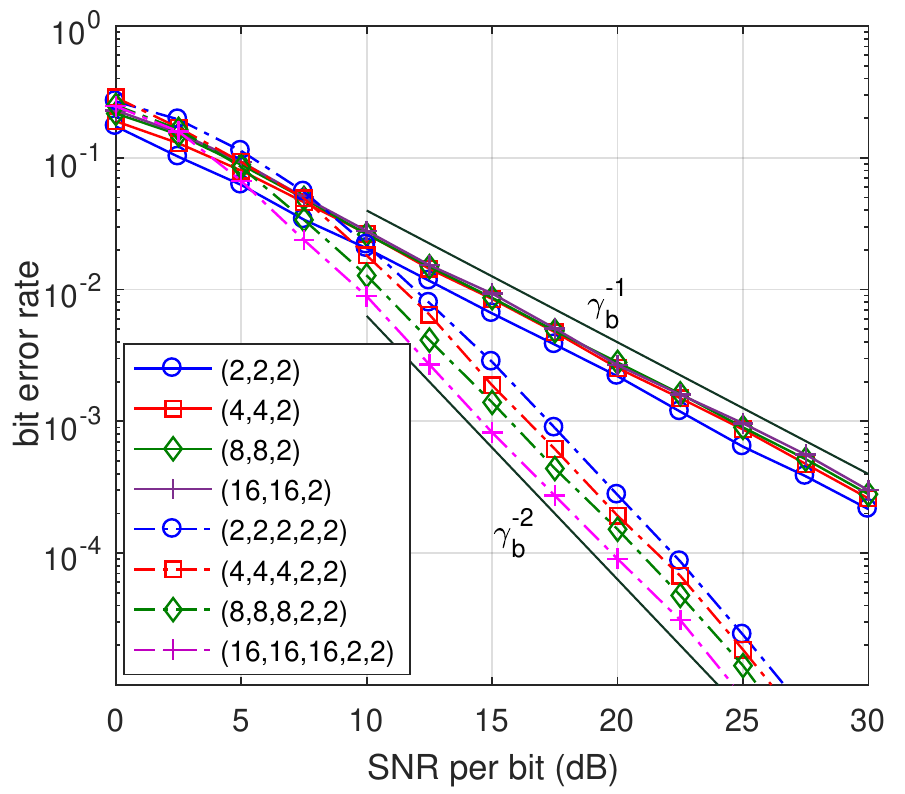}
\caption{BER performance of the proposed MD-PSM versus that of PSM for $n_T=n_{T_1}=n_{T_2} = n_R$ with QPSK, where the MD-PSM scheme achieves twice the spectral efficiency of the PSM scheme.}
\label{fig:sim_samentnr}
\end{figure}

Figure \ref{fig:sim_samentnr} depicts the performance of the proposed MD-PSM scheme versus that of the conventional PSM for $n_T=n_{T_1}=n_{T_2} = n_R$ and QPSK modulation, where the proposed scheme achieves twice the spectral efficiency of the conventional PSM scheme. Numerically, for $n_R = 2,4,8$ and $16$, the conventional PSM scheme achieves 3, 4, 5, and 6 bits/s/Hz, while the proposed MD-PSM achieves 6, 8, 10, and 12 bits/s/Hz, respectively, for the same number of receive antennas. Furthermore, the proposed scheme achieved a gain of 11.3, 12.4, 13.4, and 15 dB for $n_R=n_T=n_{T_1}=n_{T_2}$ = 2, 4, 8 and 16, respectively, at a target BER of $10^{-4}$. In addition to the gains in terms of spectral efficiency and SNR, MD-PSM achieves a higher diversity order due to its distributed nature. Based on (35) in \cite{mehana14}, the diversity order of the linear ZF precoding is $(n_T-n_R+1)$. Since two BSs are used in the MD-PSM system equipped with $n_{T_1}$ and $n_{T_2}$ transmit antennas, respectively, its total diversity order is $(n_{T_1}+n_{T_2}-2n_R+2)$, while it is $(n_T-n_R+1)$ in the case of the conventional PSM. For instance, MD-PSM scheme achieves a diversity order of 2 while PSM achieves only 1, for all the system configurations depicted in Fig. \ref{fig:sim_samentnr}. The theoretical diversity curves indicated as $\gamma_b^{-d}$, where $d$ is the diversity order and $\gamma_b$ is linear value of the SNR per bit, are also drawn in Fig. \ref{fig:sim_samentnr}. The simulation results and the diversity curves are parallel at high values of SNR, which implies that PSM and MD-PSM achieve the theoretical diversity orders indicated above. It is also shown in Fig. \ref{fig:sim_samentnr} that MD-PSM outperforms PSM for $n_T=n_{T_1}+n_{T_2}$ while achieving double its spectral efficiency. For instance, for $n_T=16$, $n_{T_1}=n_{T_2}=8$ and $n_R = 8$, MD-PSM outperforms PSM by more than 10 dB while achieving double the spectral efficiency and the diversity order. In this case, MD-PSM requires a total of 24 antennas while PSM requires 32.

\begin{figure}
\centering
\includegraphics[scale=0.82]{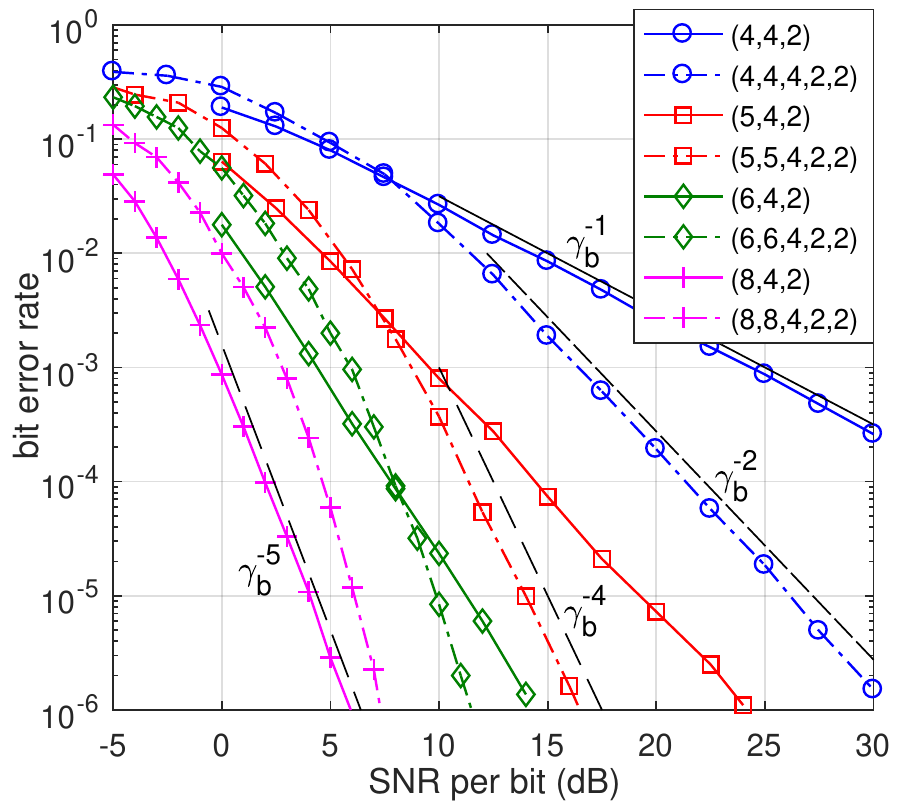}
\caption{BER performance of the proposed MD-PSM versus that of PSM for several values of $n_T=n_{T_1}=n_{T_2}$ and $n_R=4$, where the MD-PSM scheme achieves twice spectral efficiency and diversity order of the PSM scheme.}
\label{fig:sim_samenr_doublese}
\end{figure}

Figure \ref{fig:sim_samenr_doublese} depicts the BER performance of MD-PSM versus that of PSM for $n_R = 4$ and several values of $n_T=n_{T_1}=n_{T_2}$ with QPSK modulation. The proposed scheme achieves twice the spectral efficiency of the conventional PSM. The diversity curves are also depicted for several values of diversity order. At low values of SNR, the MD-PSM scheme lags the performance of the conventional PSM, mainly due to the reduction in the minimum Euclidean distance of the receive constellation set. The minimum Euclidean distance is equal to 0.515 and 1.414 in the case of MD-PSM and PSM, respectively. However, the proposed MD-PSM outperforms the PSM scheme at medium to high values of SNR, depending on the value of $n_T$, while achieving double the diversity gain. This is clearly manifested in the case of $n_T=4, 5$ and 6. The degradation in the BER performance in the case of $n_T = 8$ is tolerable, considering the gains achieved by the proposed MD-PSM in terms of spectral efficiency and diversity order as compared to the conventional PSM scheme.
\begin{figure}
\centering
\includegraphics[scale=0.82]{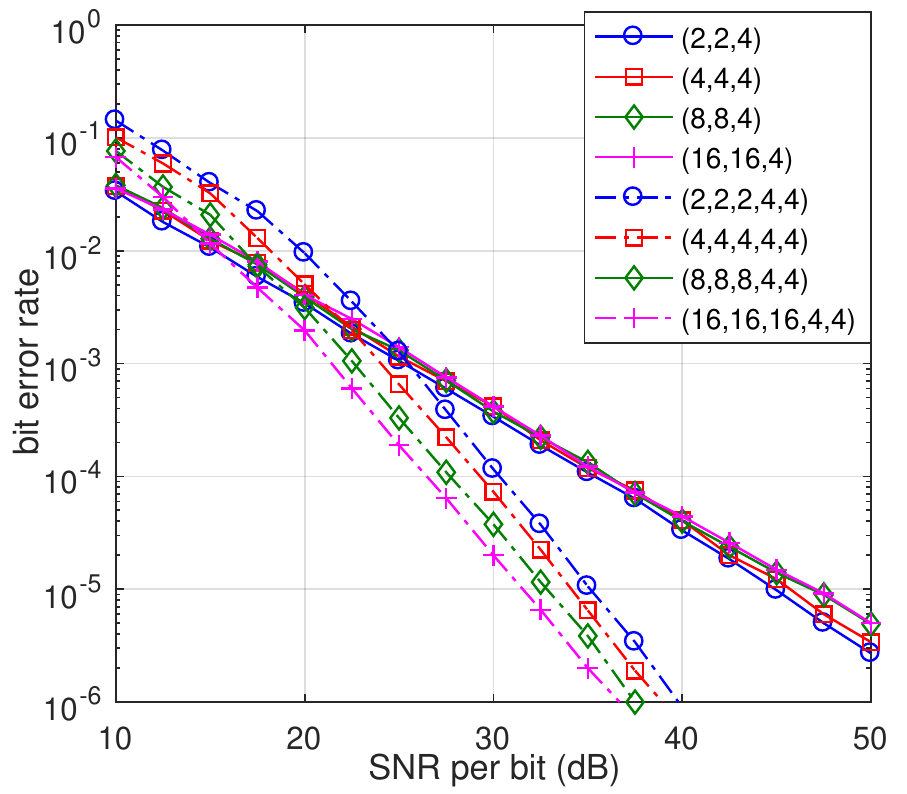}
\caption{BER performance of the proposed MD-PSM versus that of PSM for $n_T=n_{T_1}=n_{T_2} = n_R$ with 16QAM, where the MD-PSM scheme achieves twice the spectral efficiency of the PSM scheme.}
\label{fig:sim_16qamsamentnr}
\end{figure}

Figure \ref{fig:sim_16qamsamentnr} depicts the performance of the novel MD-PSM scheme versus that of PSM for $n_T=n_{T_1}=n_{T_2} = n_R$ with 16QAM. The novel MD-PSM outperforms the PSM scheme by a maximum of 16.6 dB in the case of $n_R=16$ and a minimum of 10.2 dB in the case of $n_R=2$, all at a target BER of $10^{-5}$. Besides, MD-PSM scheme doubles the diversity gain and spectral efficiency of the PSM scheme.

Finally, based on Figs. \ref{fig:angle_sim}, \ref{fig:sim_samentnr}, and \ref{fig:sim_16qamsamentnr}, for the same zero-forcing precoding and as $n_R$ increases, the BER performance of the conventional PSM deteriorates while that of the novel MD-PSM improves. As the dimension of the channel matrix increases, its minimum singular value, $\sigma_{\text{min}}$, decreases (refer to \cite{mohaisen09} and references therein). For an $n_R\times n_R$ channel matrix, Theorem 2.36 in \cite{tulino04} states that
\begin{equation}
\lim_{n_R\rightarrow \infty} {\text{Pr}}[n_R\sigma_{\text{min}} \geq x] = e^{-x-x^2/2}.
\label{eq:pr}
\end{equation}
This implies that the minimum singular value decreases as the matrix dimension increases. For instance and following (\ref{eq:pr}), ${\text{Pr}}[\sigma_{\text{min}} \geq 0.1]$ = 0.8025 and 0.0561 for $n_R=2$ and $16$, respectively. While this Theorem is sufficient to reason for the performance trend of the conventional PSM, the performance of the MD-PSM scheme, on the other hand, is also affected by the minimum Euclidean distance of the constellation set $\Omega_d$ as well as by the statistics of $\sigma_{\text{min}}$. Based on (\ref{eq:40x}), it is implied that as $n_R$ increases, the probability that $m_k \in \Omega_c$ decreases. While the minimum Euclidean distance of $\Omega_d$ is still the same regardless of the value of $n_R$, its {\it{average}} value increases as $n_R$ increases, leading to improved BER performance of the proposed MD-PSM scheme.

\vspace{10pt}
\section{\uppercase{Computational Complexity}}
\label{sec:complexity}
Based on (\ref{eq:a}), the computational complexity of the conventional PSM in terms of real multiplications is given by:
\begin{equation}
C_{{\text{PSM}}} = M(3+2n_R).
\end{equation}
Based on (\ref{eq:10})-(\ref{eq:16}), the computational complexity of the MD-PSM scheme is given by:
\begin{equation}
C_{{\text{MD-PSM}}} = (M_1+M_2+M_1M_2)(3+2n_R).
\end{equation}
For a fair comparison between the two schemes, we consider the case of equal spectral efficiency depicted in Fig. \ref{fig:sim_samese}, where MD-PSM outperforms PSM by more than 5.5 dB for all the simulated scenarios. For the two scenarios listed in Table \ref{table:comp}, MD-PSM scheme requires 37.5$\%$ and 21.9$\%$, respectively, of the computational complexity of PSM. Using the results in Table \ref{tab:1} and \ref{tab:2} leads to further reduction in the computational complexity. For instance, when one of the BSs uses 8PSk and the second uses 16QAM with $n_R=4$, the computational complexity is reduced by ($3M_1+3M_1M_2-4=404$), where the computation of the values in Table \ref{tab:2} requires four real multiplications, and the evaluation of two sine and four cosine functions. This reduces the computational complexity by 24.16$\%$.

\begin{table}
\small
\centering
\caption{Computational complexity of MD-PSM versus that of PSM for an equal spectral efficiency.}
\label{table:comp}
\def\arraystretch{1.4}

\begin{tabular}{l|c|c}
  \hline
  scheme & spectral efficiency & complexity\\
  \hline
PSM: (4, 4, 6) & \multirow{2}{*}{\begin{tabular}{@{}c@{}}8 \end{tabular}}& 704 \\
MD-PSM: (4, 4, 4, 2, 2) & & 264\\
\hline
PSM: (12,8,6) & \multirow{2}{*}{\begin{tabular}{@{}c@{}}9 \end{tabular}}& 1216 \\
MD-PSM: (12, 12, 8, 1, 2) & & 266\\
\hline
 \end{tabular}
\label{tab:3}
 \end{table}

\vspace{10pt}
\section{\uppercase{Conclusions and Future Works}}
\label{sec:conc}
In this paper, a macro-diversity PSM scheme is proposed, where two BSs communicate simultaneously with a single MS on the downlink. The proposed scheme achieves double the spectral efficiency and diversity order of the conventional PSM. Furthermore, substantial improvements in the error rate performance are achieved in several system configurations. These gains are attained through optimizing the constellation sets used at both BSs so that the minimum Euclidean distance of the receive constellation set is maximized. This is attained through rotating the constellation set used at the second BS. Taking into consideration the structure of the macro-diversity PSM scheme, a simplified receiver based on the maximum-likelihood principle is devised. Analytical formulae are also derived for parts of the maximum-likelihood's cost function for the most commonly used constellation sets so that the computational complexity is further reduced. All these gains are supported by both analytical and simulation results.

Our future work is twofold: 1) designing unconventional constellation sets, taking into consideration the structure of the proposed scheme in order to either improve the error rate performance or reduce the computational complexity, or both, and 2) investigating the combination of more sophisticated precoders with the proposed scheme.

\vspace{10pt}

\epsfysize=3.2cm
\begin{biography}{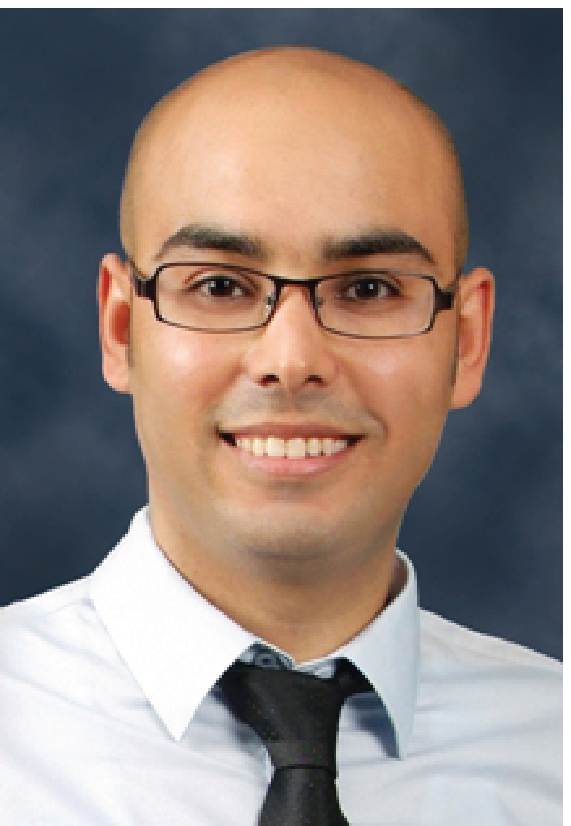}{Manar Mohaisen} received an MSc degree in communications and signal processing from the University of Nice-Sophia Antiplois, France, in 2005, and a PhD from Inha University, Rep. of Korea, in 2010, both in communications engineering. From 2001 to 2004, he was a cell planning engineer at the Palestinian Telecommunications Company. Since 2010, he has been an assistant professor at the Department of EEC Engineering, Korea Tech, Rep. of Korea. His research interests include wireless communications and social network analysis.
\end{biography}
\epsfysize=3.2cm
\begin{biography}{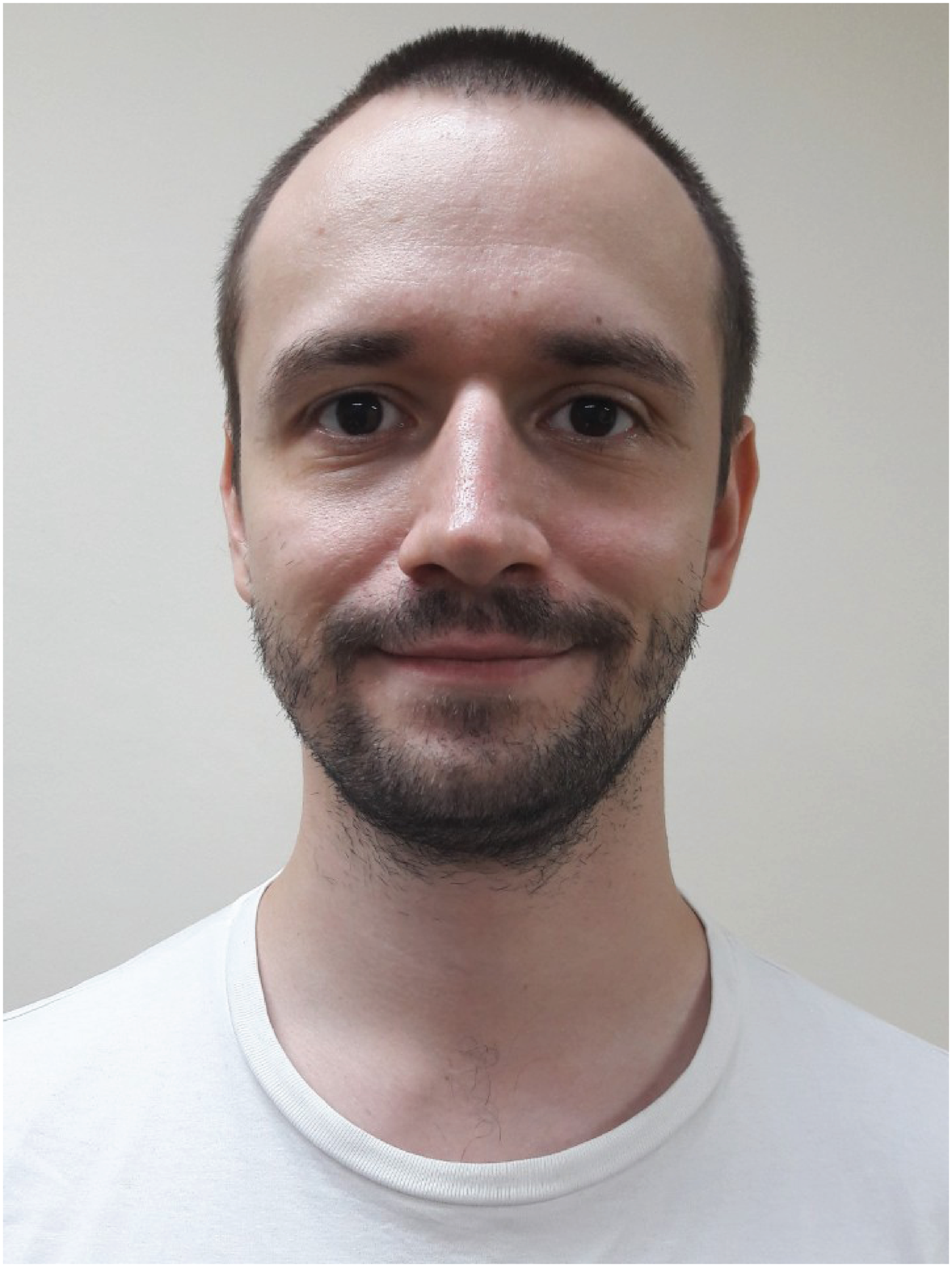}{Vitalii Pruks} received a BSc degree in computer engineering from Immanuel Kant Baltic Federal University, Russia in 2008, and MSc degree in mechanical engineering from Korea Tech, Rep. of Korea in 2016. He is currently pursuing a PhD degree at the Department of Mechanical Engineering, Korea Tech, Rep. of Korea. His research interests include computer vision, telerobotics and haptics, MIMO systems with an emphasis on spatial modulation and antenna selection.
\end{biography}


\begin{thebibliography}{99}

\bibitem{mesleh08}
R. Mesleh, H. Haas, S. Sinanovic, C. W. Ahn, and S. Yun, ``Spatial modulation,'' \emph{IEEE Transactions on Vehicular Technology}, vol. 57, no. 4, pp. 2228-2241, Jul. 2008.

\bibitem{jeg09}
J. Jeganathan, A. Ghrayeb, L. Szczecinski, and A. Ceron, ``Space shift keying modulation for MIMO channels,'' \emph{IEEE Transactions on Wireless Communications,} vol. 8, no. 7, pp. 3692-3703, July 2009 )

\bibitem{jeg08}
J. Jeganathan, A. Ghrayeb, and L. Szczecinski, ``Generalized space shift keying modulation for MIMO channels,'' in \emph{Proc. PIMRC}, pp. 1-5, 2008.

\bibitem{younis13}
A. Younis, S. Sinanovic, M. Di Renzo, R. Mesleh, and H. Haas, ``Generalized sphere decoding for spatial modulation,'' \emph{IEEE Transactions on Communications}, vol. 61, no. 7, pp. 2805-2815, July 2013.

\bibitem{younis14}
A. Younis, R. Mesleh, M. Di Renzo, and H. Haas, ``Generalized spatial modulation for large-scale MIMO,'' in \emph{Proc. Eusipco,} pp. 346-350, Sep. 2014.

\bibitem{mesleh15}
R. Mesleh, S. Ikki, and H. M. Aggoune, ``Quadrature spatial modulation,'' \emph{IEEE Transactions on Vehicular Technology}, vol. 64, no. 6, pp. 2738-2742, June 2015.

\bibitem{cheng15}
C.-C. Cheng, H. Sari, S. Sezginer, and Y. T. Su, ``Enhanced spatial modulation with multiple signal constellations,'' \emph{IEEE Transactions on Communications,} vol. 63, no. 6, pp. 2237-2248, Jun. 2015.

\bibitem{cheng16}
C.-C. Cheng, H. Sari, S. Sezginer, and Y. T. Su, ``New signal designs for enhanced spatial modulation,'' \emph{IEEE Transactions on Wireless Communications,} vol. 15, no. 11, pp. 7766-7777, Nov. 2016.

\bibitem{basar16}
E. Basar, ``Index modulation techniques for 5G wireless networks,'' \emph{IEEE Communications Magazine,} vol. 54, no. 7, Jul. 2016, pp. 168-175.

\bibitem{peel05}
C. Peel, B. Hochwald, and A. Swindlehurst, ``A vector-perturbation technique for near-capacity multiantenna multiuser communication - part I: Channel inversion and regularization,'' \emph{IEEE Transactions on Communications,} vol. 53, no. 1, Jan. 2005, pp. 195-202.

\bibitem{hochwald05}
B. Hochwald, C. Peel, and A. Swindlehurst, ``A vector-perturbation technique for near-capacity multiantenna multiuser communication - part II: Perturbation,'' \emph{IEEE Transactions on Communications,} vol. 53, no. 3, Mar. 2005, pp. 537 - 544.

\bibitem{yang11}
L.L. Yang, ``Transmitter preprocessing aided modulation for multiple-input multiple-output systems,'' in \emph{Proc. IEEE VTC}, pp. 1-5, Spring 2011.

\bibitem{mohaisen15}
M. Mohaisen, ``A Review of Fixed-Complexity Vector Perturbation for MU-MIMO,'' \emph{Journal of Information Processing Systems}, vol. 11, no. 3, pp. 354-369, 2015. 


\bibitem{tulino04}
A. Tulino and S. Verdu, \emph{Random Matrix Theory and Wireless Communications.} Delft, The Netherlands: Now, 2004.

\bibitem{telatar99}
E. Telatar, ``capacity of multi antenna gaussian channels,'' \emph{European transactions on telecommunications,} vol. 10, no. 6, pp. 585-595, 1999.


\bibitem{ge14}
X. Ge, H. Cheng, M. Guizani, abd T. Han, ``5G wireless backhaul networks: Challenges and research advances,'' \emph{IEEE Networks}, vol. 28, no. 6, pp. 6-11, Dec. 2014.

\bibitem{deberg}
M. de Berg, O. Cheong, M. van Kreveld, and M. Overmars, \emph{Computational Geometry: Algorithms and Applications, 3rd ed.} Berlin, Germany: Springer, 2008.


\bibitem{mehana14}
A. Mehana and A. Nosratinia, ``Diversity of MIMO linear precoding,'' \emph{IEEE Transactions on Information Theory,} vol. 60, no. 2, pp. 1019-1038, Feb. 2014.

\bibitem{mohaisen09}
M. Mohaisen and K.-H. Chang, ``On the achievable improvement by the linear minimum mean square error detector,'' in \emph{Proc. ISCIT 2009,} pp. 770-774, 2009.




\newpage
\end{thebibliography}
\end{document}